\documentstyle[aps]{revtex}
\begin{document}
\title{Curvature conditions for the occurrence of a class
of spacetime singularities}
\author{ Wies{\l}aw Rudnicki\thanks{E-mail:
rudnicki@atena.univ.rzeszow.pl} \, and Pawe{\l}
Zi\c{e}ba\thanks{E-mail: pzieba@atena.univ.rzeszow.pl}
\\ {\em Institute of Physics, University of Rzesz\'ow,} \\
{\em ul. Rejtana 16A, PL 35-310 Rzesz\'ow, Poland}}
\maketitle
\begin{abstract}
It has previously been shown [W. Rudnicki, Phys. Lett. A {\bf 224},
45 (1996)] that a generic gravitational collapse cannot result in
a naked singularity accompanied by closed timelike curves.
An important role in this result plays the so-called {\em
inextendibility condition}, which is required to hold for certain
incomplete null geodesics. In this paper, a theorem is proved that
establishes some relations between the inextendibility condition
and the rate of growth of the Ricci curvature along incomplete null
geodesics. This theorem shows that the inextendibility condition may
hold for a much more general class of singularities than only those of
the strong curvature type. It is also argued that some earlier cosmic
censorship results obtained for strong curvature singularities can
be extended to singularities corresponding to the inextendibility
condition.
\end{abstract}

\section{Introduction}

Recently, one of us [1] has shown that, under certain physically
reasonable conditions, a generic gravitational collapse developing from
a regular initial state cannot lead to the formation of a final state
resembling the Kerr solution with $a^{2}>m^{2}$---i.e., of a naked
singularity accompanied by closed timelike curves. This result
supports the validity of Penrose's cosmic censorship hypothesis [2]
and suggests that there may exist some deeper connection between cosmic
censorship and the chronology protection conjecture put forward by
Hawking [3]. An important role in this result plays the so-called {\em
inextendibility condition} (see Sec. II), which is assumed to be
satisfied for certain incomplete null geodesics. This condition enables
one to rule out artificial naked singularities that could easily be
created by simply removing points from otherwise well-behaved
spacetimes. The inextendibility condition is based on the idea that
physically essential singularities should always be associated with
large curvature strengths, which are in turn usually associated with
the focusing of Jacobi fields along null geodesics.

It is easily seen that the inextendibility condition will always hold
for null geodesics terminating at the so-called {\em strong curvature
singularities} defined by Tipler [4] (see below).
Singularities of this type are sometimes considered to be the {\em
only} physically reasonable singularities (cf., e.g., [5,6]).
However, strong curvature singularities can exist only if the curvature
in their neighborhood diverges strong enough [7], while it is not
unlikely that some singularities occurring in generic collapse
situations will involve a weaker divergence of the curvature. In fact,
one cannot {\em a priori} exclude the existence of some ``real''
singularities near which the curvature would remain even bounded (such
singularities occur, for example, in Taub-NUT space). Accordingly,
since we still have no fully accepted {\em necessary} condition on the
behavior of the curvature near generic singularities, one should try to
prove any cosmic censorship result under as weak a curvature condition
as possible. It would be therefore of interest, in view of the
mentioned censorship result [1], to know what are curvature conditions
for the occurrence of singularities corresponding to the
inextendibility condition. Furthermore, the inextendibility condition
has also been used in proving some other recent results [8,9] that
restrict a class of possible causality violations in classical general
relativity.

In this paper, we formulate and prove a theorem that establishes some
relations between the inextendibility condition and the rate of growth
of the Ricci curvature along incomplete null geodesics. This theorem
shows that the inextendibility condition may hold for a much more
general class of possible singularities than only those of the strong
curvature type. Our theorem will be stated in Sec. II of the paper.
In Sec. III we present a proof of the theorem; our main
mathematical tool in this proof is a Sturm-type comparison lemma for
nonoscillatory solutions of second-order differential equations. In
Sec. IV we give a few concluding remarks; in particular, we argue
that some earlier cosmic censorship results obtained for strong
curvature singularities can be extended to singularities corresponding
to the inextendibility condition.

\section{The theorem}

 To begin with, we clearly need to recall the precise
formulation of the inextendibility condition. Let $\eta(t)$ be an
affinely parametrized null geodesic, and let $Z_{1}$ and $Z_{2}$ be
two linearly independent spacelike vorticity-free Jacobi fields along
$\eta(t)$. The exterior product of these Jacobi fields defines a
spacelike area element, whose magnitude at affine parameter value $t$
we denote by $A(t)$. If we now introduce the function $z(t)$ defined by
$A(t)\equiv z^{2}(t)$, then one can show [4] that $z(t)$ satisfies the
following equation:
\begin{equation}
\frac{d^{2}z}{dt^{2}}+\frac{1}{2}(R_{ab}K^{a}K^{b}+
2\sigma^{2})z=0,
\end{equation}
where $K^{a}$ is the tangent vector to $\eta(t)$ and
$\sigma^{2}$ is a non-negative function of $t$ defined as
follows: $2\sigma^{2}\equiv
\sigma_{mn}\sigma^{mn}$ $(m,n=1,2)$. Here
$\sigma_{mn}$ is the shear tensor (see [10], p. 88) that
satisfies the equation [4]:
\begin{equation}
\frac{d}{dt}\sigma_{mn}=-C_{manb}K^{a}K^{b}-
\frac{2}{z}\frac{dz}{dt}\sigma_{mn}.
\end{equation}

In the following, by $M$ we shall denote a spacetime, i.e., a smooth,
boundaryless, connected, four-dimensional Hausdorff manifold with a
globally defined $C^{2-}$ Lorentz metric.

\vspace{3mm}
{\em Definition:  Let $\eta: (0,a]\rightarrow M$
be an affinely parametrized, incomplete null geodesic. Assume also
that $\eta(t)$ generates an achronal set, i.e., a set such that no two
points of it can be joined by a timelike curve. Then $\eta(t)$ is said
to satisfy the} {\bf inextendibility condition} {\em if for some affine
parameter value $t_{1}\in (0,a)$ there exists a solution $z(t)$ of
Eq. (1) along $\eta(t)$ such that $z(t_{1})=0$,
$dz/dt|_{t_{1}}\neq 0$ and $\lim_{t\rightarrow 0}z(t)=0$.}

\vspace{3mm}
The key idea behind the inextendibility condition is based on the fact
that any two zeros of any solution of Eq. (1), which is not identically
zero along a given null geodesic, correspond to a pair of conjugate
points along the geodesic (see [4]). From Proposition 4.5.12 of Ref.
[10] it follows that incomplete null geodesics generating achronal sets
cannot contain any pairs of conjugate points. One can thus easily
show [8] that if a geodesic $\eta: (0,a]\rightarrow M$ satisfies the
inextendibility condition, then there is $no$ extension of the
spacetime $M$, preserving all the above mentioned properties of $M$, in
which $\eta(t)$ could be extended beyond a point $\eta(0)$. This means,
according to the standard interpretation, that $\eta(t)$ should then
approach a genuine singularity of the spacetime $M$ at affine parameter
value 0. [Formally, this singularity has the same status as those
predicted by the familiar singularity theorems [10], because these
theorems predict in fact the existence of incomplete causal (usually
null) geodesics in maximally extended spacetimes satisfying just the
same topological and smoothness conditions as those imposed on $M$.]

Let us now compare the inextendibility condition with the concept of
a strong curvature singularity [4]. Consider a null geodesic $\lambda:
(0,a]\rightarrow M$ that terminates in a strong curvature singularity
at affine parameter value 0. This means that every solution $z(t)$ of
Eq. (1) along $\lambda(t)$, which vanishes for at most finitely
many points in $(0,a]$, satisfies $\lim_{t\rightarrow 0}z(t)=0$
(cf. Ref. [5], p. 160). Suppose now that $\lambda(t)$ generates an
achronal set; then any solution of Eq. (1), which is not identically
zero along $\lambda(t)$, cannot vanish for any two points in $(0,a]$ by
the argument with conjugate points mentioned above. Thus, for {\em all}
$t_{1}\in (0,a]$ and for {\em all} solutions $z(t)$ of Eq. (1) along
$\lambda(t)$ with initial conditions $z(t_{1})=0$ we will have
$\lim_{t\rightarrow 0}z(t)=0$. It is thus clear that any null geodesic
terminating in Tipler's strong curvature singularity and generating an
achronal set must always satisfy the inextendibility condition. Notice
also that the terms ``all'' emphasized above imply, via Eqs. (1) and
(2), that $\lambda(t)$ can terminate in the strong curvature
singularity only if the curvature diverges strong enough along
$\lambda(t)$ as $t\rightarrow 0$, while the inextendibility condition
could actually be satisfied for $\lambda(t)$ even if the curvature
along it would remain bounded. Indeed, the theorem stated below makes
it clear [see condition (i)] that the curvature need not necessarily
diverge along geodesics satisfying the inextendibility condition.

\vspace{3mm}
{\bf Theorem:} \, {\em Let $\eta: (0,a]\rightarrow M$ be an affinely
parametrized, incomplete null geodesic generating an achronal set.
Suppose that the Ricci tensor term $r(t)\equiv R_{ab}K^{a}K^{b}$ along
$\eta(t)$, where $t$ is the affine parameter and $K^{a}$ is the tangent
vector to $\eta(t)$, obeys at least one of the following conditions.} \\
 (i) {\em There exists an affine parameter value $b\in (0,a)$ such that
 $\inf\{r(t)| \, 0<t<b\}\geq 2(\pi/b)^{2}$.} \\
(ii) {\em There exist an affine parameter value $c\in (0,a)$ and a
constant $\mu\in (0,2)$ such that
$r(t)\geq \kappa t^{-\mu}$ for all $t\in (0,c]$,

where $\kappa = (2/3)(33-26\mu+5\mu^{2})c^{\mu-2}$.}

{\em Then $\eta(t)$ satisfies the inextendibility condition. }

\vspace{3mm}
{\em Remark 1:} From the proof of this theorem, which is given below,
it may be seen that the parameter values $b$ and $c$ mentioned
above in conditions (i) and (ii) correspond to the parameter value
$t_{1}$ occurring in the definition of the inextendibility condition.

{\em Remark 2:} Since in the theorem $\eta(t)$ is assumed to be a
generator of an achronal set, $\eta(t)$ cannot contain any pair of
conjugate points, and so one can expect that there should exist an {\em
upper} limit on the rate of growth of the curvature along $\eta(t)$.
Indeed, from Theorems (3) and (4) of Ref. [11] it follows immediately
that the Ricci tensor term $r(t)$ along $\eta(t)$ must satisfy the
following two conditions: (1) there is no affine parameter value $b'\in
(0,a]$ such that $\inf\{r(t)|\, 0<t<b'\}>8(\pi/b')^{2}$; and (2) if
$r(t)\geq 0$ on $\eta(t)$, then $\lim_{t\rightarrow 0}\inf
t^{2}r(t)\leq 1/2$. Similar restrictions on the growth of the Weyl part
of the curvature along $\eta(t)$ can be obtained from Proposition 2.2
of Ref. [12].

In the context of our theorem, it is worth recalling the analogous
results obtained by Clarke and Kr\'olak [7] for singularities of the
strong curvature type. They have been obtained for two definitions of a
strong curvature singularity: the original one formulated by
Tipler [4] and its modification proposed by Kr\'olak [6]. According
to these results, if a null geodesic $\eta: (0,a]\rightarrow M$
terminates at affine parameter value 0 in a strong curvature
singularity defined by Tipler (resp., by Kr\'olak), then there must
exist some affine parameter value $c\in (0,a]$ such that
$R_{ab}K^{a}K^{b}>A t^{-2}$ (resp., $R_{ab}K^{a}K^{b}>A t^{-1}$) on
$(0,c]$, where $K^{a}$ is the tangent vector to $\eta(t)$, $t$ is the
affine parameter, and $A$ is some fixed positive constant. [Or very
similar conditions on the rate of growth of the Weyl part of the
curvature along $\eta(t)$ must be satisfied; see Corollary 2 of Ref.
7.] Comparing these results with condition (ii) of our theorem we see
that singularities of the strong curvature type involve a considerably
stronger divergence of the Ricci tensor term $R_{ab}K^{a}K^{b}$ than
singularities corresponding to the inextendibility condition. There may
thus exist a large class of curvature singularities that are not
strong in the sense of the definition of Tipler or Kr\'olak, but they
may still satisfy the inextendibility condition. Note also that the
above conditions for strong curvature singularities are the {\em
necessary} ones, whereas conditions (i) and (ii) of our theorem are
only {\em sufficient} to ensure that the inextendibility condition does
hold for a given geodesic. This implies that the inextendibility
condition might be satisfied in more general situations than only those
characterized by conditions (i) and (ii).

\section{Proof of the theorem}

\vspace{4mm}
Now we shall prove the theorem; our main tool in this
proof will be the following comparison lemma.

\vspace{3mm}
{\em Lemma (The comparison lemma):  Suppose that $u(s)$ is a solution
of the equation
\begin{displaymath}
\frac{d^{2}u}{ds^{2}}+F(s)u(s)=0
\end{displaymath}
on an interval $(a,b]$ with initial conditions:
$u(b)=0$ and $du/ds|_{b}\neq 0$. Let $v(s)$ be a solution of
\begin{displaymath}
\frac{d^{2}v}{ds^{2}}+G(s)v(s)=0
\end{displaymath}
on $(a,b]$ such that $v(b)=0$,
$dv/ds|_{b}=du/ds|_{b}$ and $v(s)>0$ on $(a,b)$.
Assume also that $F(s)$ and $G(s)$ are piecewise continuous on
$(a,b]$, and let $G(s)\geq F(s)$ on $(a,b]$. Then
$u(s)\geq v(s)$ on $(a,b]$. }

\vspace{3mm}
$Proof:$ \, The proof of this lemma is based essentially on Theorem 1.2
of Ref. [13], p. 210. To apply this theorem in its original form, it is
convenient to reparametrize both of the equations in the lemma
introducing the parameter $t=-s$ instead of $s$. Note that this
reparametrization does not change the form of the equations. Clearly,
we shall now have established the lemma if we show that for any $c\in
(a,b)$, $u(t)\geq v(t)$ on $[-b,-c]$.

Consider the ratio $u(t)/v(t)$. Since $v(t)>0$ on $(-b,-a)$, it is
well defined on $(-b,-c]$. Using l'Hospital's rule, we get
\begin{displaymath}
\lim\limits_{t\rightarrow -b}\frac{u(t)}{v(t)}=1.
\end{displaymath}
Therefore, as $v(t)>0$ on $(-b,-c]$, to show that
$u(t)\geq v(t)$ on $[-b,-c]$, it suffices to show that
\begin{displaymath}
\frac{d}{dt}\left[\frac{u(t)}{v(t)}\right]\geq 0
\end{displaymath}
on $(-b,-c]$. It is easy to see that this inequality
holds if
\begin{equation}
\frac{v(t)}{\dot{v}(t)}\geq \frac{u(t)}{\dot{u}(t)}
\end{equation}
on $(-b,-c]$, where the overdot denotes the first derivative with
respect to $t$. Since $F(t)$ and $G(t)$ are piecewise continuous on
$[-b,-c]$, by Theorem 1.2 of Ref. 13, p. 210, we have
\begin{displaymath}
\tan^{-1}\left[\frac{v(t)}{\dot{v}(t)}\right]\geq
\tan^{-1}\left[\frac{u(t)}{\dot{u}(t)} \right]
\end{displaymath}
for all $t\in [-b,-c]$. Thus, as $\tan^{-1}$ is an increasing
function, the inequality (3) does hold as it is desirable.
\hfill $\Box$

{\em Proof of the theorem:}

(Part I) Suppose the condition (i) is satisfied. Let $z_{0}(t)$ be a
solution of Eq. (1) along $\eta(t)$ such that $z_{0}(t)$ is
not identically zero on $(0,b]$ and $z_{0}(b)=0$, where $b$ is the
parameter value mentioned in condition (i). Clearly, such a solution
will always exist. Since $\eta(t)$ generates an achronal set,
$z_{0}(t)$ can vanish nowhere in $(0,b)$; otherwise $\eta(t)$ would
have a pair of conjugate points in $(0,b]$ (see Ref. [4]), which would
contradict, by Proposition 4.5.12 of Ref. [10], the achronality of
$\eta(t)$. Notice also that Eq. (1) is linear, and so the function
$-z_{0}(t)$ will be a solution of Eq. (1) as well. Thus, as $z_{0}(t)
\neq 0$ on $(0,b)$, without loss of generality we can assume that
$z_{0}(t)>0$ on $(0,b)$. This implies, as $z_{0}(b)=0$, that
$dz_{0}/dt|_{b}\leq 0$. Since $z_{0}(t)>0$ on $(0,b)$, and condition
(i) holds, from Eq. (1) we see at once that $z_{0}(t)$ must be a
concave function on $(0,b]$. This makes it obvious that
$dz_{0}/dt|_{b}\neq 0$, and so we must have $dz_{0}/dt|_{b}=\alpha<0$.
Let us now define the function $z_{1}(t)\equiv-(1/\alpha)z_{0}(t)$. As
Eq. (1) is linear, it is clear that $z_{1}(t)$ will be a solution of
Eq. (1) along $\eta(t)$; notice also that $z_{1}(t)>0$ on $(0,b)$,
$z_{1}(b)=0$ and $dz_{1}/dt|_{b}=-1$.

Consider now the equation
\begin{equation}
\frac{d^{2}x}{dt^{2}}+\omega x(t)=0,
\end{equation}
where $\omega=\frac{1}{2}\inf\{r(t)|\, 0<t\leq b\}$ and $r(t)$ is the
function defined in the theorem. Notice that $\omega>0$ by condition
(i). Let $x_{1}(t)$ be a solution of Eq. (4) on $(0,b]$ with initial
conditions: $x_{1}(b)=0$ and $dx_{1}/dt|_{b}=-1$. It is a simple matter
to see that $x_{1}(t)=\omega^{-1/2}\sin[\omega^{1/2} (b-t)]$. Let us
now apply the comparison lemma to the equations (1) and (4) and their
solutions $z_{1}(t)$ and $x_{1}(t)$. Since $\omega\leq\frac{1}{2}r(t)$
on $(0,b]$, by the comparison lemma we must have
$x_{1}(t)\geq z_{1}(t)$ on $(0,b]$. Consequently, as $z_{1}(t)>0$ on
$(0,b)$, we obtain $x_{1}(t)>0$ on $(0,b)$. This implies, by the
above form of $x_{1}(t)$, that $\omega\leq(\pi/b)^{2}$. But $\omega\geq
(\pi/b)^{2}$ by condition (1). We must thus have $\omega=(\pi/b)^{2}$,
which gives $\lim_{t\rightarrow 0}x_{1}(t)=0$. Therefore
$\lim_{t\rightarrow 0}z_{1}(t)=0$ since $x_{1}(t)\geq z_{1}(t)> 0$ on
$(0,b)$. This means that $\eta(t)$ does satisfy the inextendibility
condition.

(Part II) The task is now to prove the theorem in the case when
condition (ii) holds. For this purpose, let us consider the following
equation
\begin{equation}
\frac{d^{2}y}{dt^{2}}+Bt^{-\mu}y(t)=0
\end{equation}
on $(0,c]$, where $B=\kappa /2$, and $\kappa$, $\mu$ and $c$ are some
fixed constants mentioned in the condition (ii). Let $y_{1}(t)$ be a
solution of this equation with initial conditions: $y_{1}(c)=0$ and
$dy_{1}/dt|_{c}=-1$. Let $z_{2}(t)$ be a solution of Eq. (1) along
$\eta(t)$ such that $z_{2}(c)=0$ and $dz_{2}/dt|_{c}=-1$. [There is no
loss of generality in assuming $z_{2}(t)$ to exist; the existence of
$z_{2}(t)$ can be established in the same manner as the existence of
the analogous solution $z_{1}(t)$ considered in the first part of the
proof.] Clearly, the solution $z_{2}(t)$, just as $z_{1}(t)$, can vanish
nowhere in $(0,c)$ by the argument with conjugate points. Therefore, as
$dz_{2}/dt|_{c}=-1$, we must have $z_{2}(t)>0$ on $(0,c)$. Let us now
apply the comparison lemma to the equations (1) and (5) and their
solutions $z_{2}(t)$ and $y_{1}(t)$. By condition (ii) we
have $r(t) \geq \kappa
t^{-\mu}$ on $(0,c]$. Thus by the comparison lemma, we must have
$y_{1}(t)\geq z_{2}(t)$ on $(0,c]$. Of course, in order to prove the
theorem, it suffices to show that $\lim_{t\rightarrow 0}z_{2}(t)=0$.
Thus, as $y_{1}(t)\geq z_{2}(t)>0$ on $(0,c)$, to complete the proof it
suffices to show that $\lim_{t\rightarrow 0}y_{1}(t)=0$. We shall show
below that $y_{1}(t)$ does possess this property.

To this end, let us first find the general solution of Eq. (5). It is
easy to check that if one puts $x=t$, $\alpha=1/2$, $\beta=
2\sqrt{B}(2-\mu)^{-1}$, $\gamma=(2-\mu)/2$ and $n=(2-\mu)^{-1}$
into the equation (4.1) of Ref. [14], p. 138, then this equation reduces
to our equation (5). Thus, according to the solution (4.3) of Eq. (4.1)
of Ref. [14], our equation (5) has the following general solution
\begin{equation}
y(t)=t^{1/2}[C_{1}J_{n}(\beta t^{\gamma})+C_{2}Y_{n}(\beta t^{\gamma})],
\end{equation}
where $C_{1}$ and $C_{2}$ are arbitrary constants of integration, and
$J_{n}(\beta t^{\gamma})$ and $Y_{n}(\beta t^{\gamma})$ are the Bessel
functions of order $n$, of the first and second kind, respectively.
Since $\mu\in (0,2)$, from the above relations it follows that
$1/2<n<\infty$, $\sqrt{B}<\beta<\infty$ and $0<\gamma<1$.

Let us recall that any Bessel function of the first kind has infinitely
many positive zeros (cf., e.g., [15], p. 29). Let $j_{n,1}$ be the
first positive zero of the function $J_{n}(\beta t^{\gamma})$, i.e.,
$J_{n}(j_{n,1})=0$ and $J_{n}(\beta t^{\gamma})\neq 0$ as long as
$0<\beta t^{\gamma}<j_{n,1}$. Since $n>1/2$, $j_{n,1}$ must satisfy the
following relation (see Eq. (2) of Ref. [15], p. 29):
\begin{equation}
j_{n,1}<2[(n+1)(n+5)/3]^{1/2}.
\end{equation}
For $J_{n}(\beta t^{\gamma})$ we now define $L$ to be the number such
that $j_{n,1}=L\beta c^{\gamma}$. Putting this into (7), and taking
into account the fact that $\beta =(2\kappa)^{1/2}(2-\mu)^{-1}$,
$\kappa=3^{-1}(66-52\mu+10\mu^{2})c^{\mu-2}$, $\gamma=(2-\mu)/2$ and
$n=(2-\mu)^{-1}$, we readily find that $L^{2}<1$.

Consider now equation (5) with $B$ replaced by $B'=L^{2} B$.
 Let $y_{2}(t)$ be a solution of this equation on $(0,c]$
with initial conditions: $y_{2}(c)=0$ and $dy_{2}/dt|_{c}=-1$. The
general form of this solution is given by (6), where $\beta$ should be
replaced by $\beta'= 2\sqrt{B'}(2-\mu)^{-1}$ (notice that $\beta'=
L\beta$). Let us now insert the initial conditions for $y_{2}(t)$ into
this general solution in order to determine for $y_{2}(t)$ the
constants $C_{1}$ and $C_{2}$ occurring in (6). To find the first
derivative of the general solution (6), we use the following recurrence
formula
\begin{displaymath}
\frac{dJ_{n}(x)}{dx}=-J_{n+1}(x)+\frac{n}{x}J_{n}(x),
\end{displaymath}
which is also valid for $Y_{n}(x)$ (see [15], p. 197). We can now
easily calculate the constants $C_{1}$ and $C_{2}$; the result is as
follows
\begin{equation}
C_{1}=\frac{Y_{n}(\beta'c^{\gamma})}{\beta'\gamma c^{\gamma
-1/2}\left[Y_{n}(\beta'c^{\gamma})
J_{n+1}(\beta'c^{\gamma})-Y_{n+1}(\beta'c^{\gamma})
J_{n}(\beta'c^{\gamma})\right]}
\end{equation}
and
\begin{equation}
C_{2}=\frac{-J_{n}(\beta'c^{\gamma})}{\beta'\gamma c^{\gamma
-1/2}\left[Y_{n}(\beta' c^{\gamma}) J_{n+1}(\beta'
c^{\gamma})-Y_{n+1}(\beta' c^{\gamma})J_{n}(\beta' c^{\gamma})\right]}.
\end{equation}
As $\beta'=L \beta$, from the above definition of $L$ it is clear
that $\beta' c^{\gamma}=j_{n,1}$. Thus $J_{n}(\beta' c^{\gamma}) =0$
and the numerator in (9) must vanish. As $J_{n}(\beta' c^{\gamma})=0$,
the denominator in (9) can vanish only if $Y_{n}(\beta'c^{\gamma})
J_{n+1}(\beta' c^{\gamma})=0$. But the Bessel functions $J_{n+1}$ and
$Y_{n}$ cannot have any common zeros with the Bessel function $J_{n}$
(see [15], pp. 29-32), and so the denominator in (9) cannot vanish.
We thus have $C_{2}=0$ and, by (6) and (8), the solution $y_{2}(t)$ can
be written as follows
\begin{equation}
y_{2}(t)=C_{1}t^{1/2}J_{n}(\beta' t^{\gamma}),
\end{equation}
where $C_{1}=[\beta'\gamma c^{\gamma -1/2}J_{n+1}(\beta'
c^{\gamma})]^{-1}$.

Let us now compare the solutions $y_{1}(t)$ and $y_{2}(t)$ by means of
the comparison lemma. Recall that $y_{1}(t)$ is a solution of equation
(5) with $B=\kappa/2$, while $y_{2}(t)$ is a solution of the same
equation with $B$ replaced by $B'=L^{2}\kappa/2$. Since $L^{2}<1$, by
the comparison lemma we must have $y_{2}(t)\geq y_{1}(t)$ for all $t\in
(0,c]$. We recall that any Bessel function $J_{k}(x)$ of the first kind
with real $x$ and $k>0$ is continuous at $x=0$ (cf. [15], p. 182).
Thus, as $n>1/2$ and $0<\gamma<1$, from (10) it follows immediately
that $\lim_{t\rightarrow 0}y_{2}(t)=0$. Therefore, as $y_{2}(t)\geq
y_{1}(t)>0$ on $(0,c)$, we obtain $\lim_{t\rightarrow 0}y_{1}(t)=0$,
which completes the proof. \hfill $\Box$

\section{Concluding remarks}

We have been concerned in this paper with the problem of determining
what are curvature conditions for the occurrence of singularities
corresponding to the inextendibility condition. We have found two such
sufficient conditions concerning the behavior of the Ricci tensor term
$R_{ab}K^{a}K^{b}$ along incomplete null geodesics---these are
conditions (i) and (ii) of the theorem stated in Sec. II. This
theorem shows that the inextendibility condition may hold for a
considerably larger class of possible singularities than only those of
the strong curvature type. In particular, condition (i) of the theorem
shows that the inextendibility condition may hold even if the curvature
along incomplete geodesics would remain bounded. In this context, it is
worth recalling that singularities predicted by the famous singularity
theorems [10] can be interpreted as regions of the universe at which
the normal classical spacetime picture and/or certain energy conditions
break down, and this may occur in regions where the curvature, though
extremely large, still remains finite. Accordingly, if one attempts to
establish, for example, whether or not these singular regions will
conform to any cosmic censorship principle, it would be well to try to
characterize, if necessary, incomplete geodesics terminating in these
regions by a condition that may hold even if the curvature along the
geodesics would remain bounded. One possible candidate for such a
condition may thus be the inextendibility condition.

It should also be stressed here that some earlier cosmic censorship
theorems [6,16,17] proved for strong curvature singularities can
be extended to singularities corresponding to the inextendibility
condition. To see this, let us first recall that these theorems show,
briefly, that under certain restrictions imposed on the causal
structure, strong curvature singularities are censored (see Refs.
[6,16,17] for details). Proofs of these theorems are, in essence,
alike. In a brief outline, they run as follows. First, one shows that
if the theorem under consideration were false, then there would have to
exist a sequence $\{\mu_{i}\}$ of future endless, future complete null
geodesics converging to a null geodesic $\mu$ that terminates in the
future at a strong curvature singularity. One also shows that $\mu$ and
all the $\mu_{i}$ must be generators of achronal sets. As all $\mu_{i}$
are achronal, none of them can have a pair of conjugate points, and so
any irrotational congruence of Jacobi fields along any $\mu_{i}$ cannot
be refocused. As $\{\mu_{i}\}$ converges to $\mu$, this must then
imply, by continuity, that any irrotational congruence of Jacobi fields
along $\mu$ cannot be refocused as well. However, as $\mu$ terminates
in a strong curvature singularity, all irrotational congruences of
Jacobi fields along $\mu$ should be refocused. This gives the required
contradiction. It is not difficult to see, however, that this
contradiction can equally well be obtained if $\mu$ would be assumed to
satisfy the inextendibility condition, for this condition holds if at
least one irrotational congruence of Jacobi fields along a given
geodesic is refocused. It is thus clear that the censorship theorems
given in Refs. [6,16,17] are unnecessarily restricted to strong
curvature singularities and they can be extended to singularities
corresponding to the inextendibility condition.

\acknowledgments
This research was supported in part by the Polish State
Committee for Scientific Research (KBN) under Grant No. 2 P03B 073 15.

\end{document}